\documentclass[twocolumn,amssymb,prb]{revtex4}

\usepackage{graphicx}
\usepackage{dcolumn}
\usepackage{bm}

\begin{document}

\title{Genetic algorithms for the numerical solution of \\variational problems without analytic trial functions}

\author{Carlos D. Toledo}
\affiliation{Center for Intelligent Systems, Monterrey Institute of Technology\\ Campus Monterrey, Monterrey, Nuevo Le\'on 64849, M\'exico
}

\date{\today}

\begin{abstract}
A coding of functions that allows a genetic algorithm to minimize functionals without analytic trial functions is presented and implemented for solving numerically some instances of variational problems from physics.  
\end{abstract}

\maketitle

\section{INTRODUCTION}
The \emph{genetic algorithm} (GA)\cite{Gold1,Gold2,Whitley} has become popular as as a reliable computerized method for solving pro\-blems from a wide range of domains, such as function optimization, handling them even in nonlinear, multidimensional search spaces. A conventional GA is a stochastic search method inspired in natural selection and the darwinian idea of the survival of the fittest in which the possible solutions of a problem (metaphorically the points in the search space) are coded as fixed-length strings of characters of an alphabet (usually binary) that resemble the chromosomes of alive beings. A GA evolves a population of search ``points'' chosen randomly appliying iteratively on them operators of \emph{selection}, \emph{crossover} and \emph{mutation} for creating new populations (generations).

Selection consists in giving a proportionally bigger number of offspring to the fitter individuals so the characteristics that make them better prevail. The combination of this characteristics for generating new individuals is achieved through crossover, that is the interchange of portions of the strings of characters of two individuals paired randomly, giving birth to two new individuals for the next generation. In its simplest form in a GA all individuals are removed (die) after reproduction. 

The last iterative step consists in making random changes to the strings of individuals chosen with a small probability, which is named mutation after the natural process that it resembles. After some generations the individuals tend to concentrate around the fittest ``points'' in the search space, so it can be said that all of the process was a way of optimizing the function employed to determine the fitting.

The predominant kind of optimization problems attacked with GAs have been those in which the strings code literally points in a  multidimensional space, where each dimension represents a parameter or variable of interest. When the potential solutions of a problem are \emph{functions} (as it is the case for variational problems\cite{Arfken}), not points, the most popular GA approach developed to date has been that of choosing a set of analytic trial functions and combining them in the fittest way. There are two main ways for doing so: weightening them, case where a string of weights is an individual, or using \emph{genetic programming} (GP)\cite{Koza} where the trial functions and the mathematical operators needed for combining them are the alphabet that gives shape to each member of the population.

In this paper a way to directly represent numerical functions as strings (individuals) of a GA is presented, followed by its successful implementation on some instances of variational problems from physics.    
    
\section{ANGULAR GENES ARE NOT REAL GENES}
Lets take $\mathfrak{G}$ as the alphabet chosen to code individuals. In the literature of GAs (without taking GP into account) two main alphabets are usually discussed, the so called \emph{binary genes} $\mathfrak{G}={0,1}$ and the \emph{real genes} $\mathfrak{G}=\Re$. Even when there is not any special restriction on $\mathfrak{G}$ in the definition of the GA, only the implicit warning that it must facilitate the heredity of the fittest characteristics for obtaining acceptable results, it is curious how the attention in the field has been biased toward the binary and real alphabets. One of the goals of this paper is to emphazise the importance of focusing attention in other alphabets, in the extra information that is possible to get from them, specifically in an \emph{angular} one, which will be called from now on \emph{angular genes}. Aren't angular genes just real genes? The distinction made is based in the commonly forgotten fact that angles are not numbers,\cite{Lakoff} they are an entity by themselves.

The angular genes code piecewise functions as a string $\alpha$ of the angles between each consecutive pair of linear segments. For any combination of angles it is possible to scale and rotate the collection to fit the initial and final desired values. Taking $y_1=y(x_1)$ and $y_N=y(x_N)$ as the initial and final values of the piecewise function $y_k=y(x_k)$  to be represented in the range $(x_1,x_N)$ with $N-1$ linear segments and $k=2,3,\ldots,N$, the coding is defined as follows:

\[R = \sqrt {(y_N  - y_1 )^2  + (x_N  - x_1 )^2 }\]
\[\beta = \tan ^{-1}[(y_N  - y_1 )/(x_N  - x_1 )]\]

\[
r^2  = \left[ {\sum\limits_{i = 2}^N {s_{i-1} \cos \left( { \sum\limits_{j = 1}^{i-1} {\alpha_j } } \right)} } \right]^2  + \left[ {\sum\limits_{i = 2}^N {s_{i-1} \sin \left( { \sum\limits_{j = 1}^{i-1} {\alpha_j } } \right)} } \right]^2 
\]

\[
\gamma  = \tan ^{ - 1} \left[ {{\raise0.7ex\hbox{${\sum\limits_{i = 2}^N {s_{i-1} \cos \left( { \sum\limits_{j = 1}^{i-1} {\alpha_j } } \right)} }$} \!\mathord{\left/
 {\vphantom {{\sum\limits_{i = 2}^N {s_{i-1} \cos \left( { \sum\limits_{j = 1}^{i-1} {\alpha_j } } \right)} } {\sum\limits_{i = 2}^N {s_{i-1} \sin \left( { \sum\limits_{j = 1}^{i-1} {\alpha_j } } \right)} }}}\right.\kern-\nulldelimiterspace}
\!\lower0.7ex\hbox{${\sum\limits_{i = 2}^N {s_{i-1} \sin \left( { \sum\limits_{j = 1}^{i-1} {\alpha_j } } \right)} }$}}} \right]
\]

\begin{equation} \label{eq:x}
x_k  = \sum\limits_{i = 2}^k {\left( {{\raise0.7ex\hbox{$R$} \!\mathord{\left/
 {\vphantom {R r}}\right.\kern-\nulldelimiterspace}
\!\lower0.7ex\hbox{$r$}}} \right)s_{i-1} \cos \left( { \sum\limits_{j = 1}^{i-1} {\alpha_j  - \gamma  + \beta } } \right)}+x_1 
\end{equation}

\begin{equation}\label{eq:y}
y_k  = \sum\limits_{i = 2}^k {\left( {{\raise0.7ex\hbox{$R$} \!\mathord{\left/
 {\vphantom {R r}}\right.\kern-\nulldelimiterspace}
\!\lower0.7ex\hbox{$r$}}} \right)s_{i-1} \sin \left( { \sum\limits_{j = 1}^{i-1} {\alpha_j  - \gamma  + \beta } } \right)}+y_1 
\end{equation}

Where $-\sigma \leq \alpha_j \leq \sigma$ and $s$ is a string of real numbers $0 < s_i \leq 1$ that codes the relative length of each linear segment and together with $\alpha$ forms an individual. Having (\ref{eq:x}) and (\ref{eq:y}) we can further define:

\[
\Delta x_i  = \left( {{\raise0.7ex\hbox{$R$} \!\mathord{\left/
 {\vphantom {R r}}\right.\kern-\nulldelimiterspace}
\!\lower0.7ex\hbox{$r$}}} \right)s_i \cos \left( {\sum\limits_{j = 1}^i {\alpha_j  - \gamma  + \beta } } \right)
\]
\[ \Delta y_i  = \left( {{\raise0.7ex\hbox{$R$} \!\mathord{\left/
 {\vphantom {R r}}\right.\kern-\nulldelimiterspace}
\!\lower0.7ex\hbox{$r$}}} \right)s_i \sin \left( {\sum\limits_{j = 1}^i {\alpha_j  - \gamma  + \beta } } \right)
\]

\begin{equation}
y_i ' = \tan \left( {\sum\limits_{j = 1}^i {\alpha_j  - \gamma  + \beta } } \right)
\end{equation}

\begin{equation}
y_i '' = \tan (\alpha_{i+1}) (1 + y_i 'y_{i + 1} ')/\Delta x_i 
\end{equation}

Taking $\Theta_i=\Theta(x_i)$ as the evaluation in $x_i$ of the function that minimizes the functional, for the best found individual we have:

\begin{equation}\label{eq:e}
\left| {\alpha _{i + 1}  - \tan ^{ - 1} \left( {\frac{{\Theta _{i + 1} ' - \Theta _i '}}{{1 + \Theta _i '\Theta _{i + 1} '}}} \right)} \right| < \varepsilon 
\end{equation}

Equation (\ref{eq:e}) is a measure of the error of the best approximation found that clarifies the influence of a proper choice of $\sigma$ according to the problem. If $\sigma$ is too small the error can be surely surpassed but if it is too big the search space grows. 

Ignoring the differences in a tenth of $\sigma$ and $1$ the search space explored has a size of approximately $10^{N-1}\times20^{N-1}$. 

\section{EXAMPLES}
The presented coding was used to solve instances of four well known variational problems from physics. In each case the population used had a size of 100 with $N=101$, probability of mutation of $0.05$ and $500$ generations. The number of runs made for all cases was ten. For  the three first cases it was not needed the help of $s$, so $s_i=1$, but not for the last where $0<s_i\leq1$. The algorithm was written in MATLAB and implemented in a personal computer with Pentium(R)4 CPU, 2.4GHz, 448 MB RAM.

\subsection{Curve of shortest distance in the Euclidian plane}
The functional to minimize is
\[
J = \int_{x_1 ,y_1 }^{x_N ,y_N } {\sqrt {(dx)^2  + (dy)^2 } } 
\]
whose known solution\cite{Arfken} is the straight line $y=ax+b$. For the case $x_1=y_1=y_N=0,\: x_N=1$ with minimum $J=1$, the average solution found by the algorithm with $\sigma=0.005\pi$ was $J=1+1.66\times10^-5$ with standard deviation  of $1.2\times10^-6$.

\subsection{Curve of minimum revolution area}
Considering two paralel coaxial wire circles to be connected by a surface of minimum area that is generated by revolving a curve $y(x)$ about the $x$-axis, the functional to minimize is

\[
J = \int_{x_1 ,y_1 }^{x_N ,y_N } {2\pi y\sqrt {(dx)^2  + (dy)^2 }} 
\]
whose known solution\cite{Arfken} is the catenoid $y=\cosh(ax+b)/a$. For the case $-x_1=x_N=0.5,\: y_1=y_N=1$ with minimum $J=5.9917$, the average solution found by the algorithm with $\sigma=0.005\pi$ was $J=5.9919$ with standard deviation  of $1.4\times10^-5$.

\subsection{Fermat's principle}
According to Fermat's principle light will follow the path $y(x)$ for which
\[
J = \int_{x_1 ,y_1 }^{x_N ,y_N } {n(x,y)\sqrt {(dx)^2  + (dy)^2 }} 
\]
is minimum when $n$ is the index of refraction. When $n=e^y$ the solution is $y=\ln(a/\cos(x+b))$. For the case $-x_1=x_N=1,\: y_1=y_N=1$ with minimum $J=4.5749$, the average solution found by the algorithm with $\sigma=0.01\pi$ was $J=4.5752$ with standard deviation  of $5.2\times10^-5$.

\subsection{The energies of the hydrogen atom}
The hydrogen atom\cite{Cohen} is the quantum system made of a proton and an electron whose energies, without taking into account the degeneracies, can be found minimizing the functional
\[
E_{n}  = \frac{1}{c}\int_0^\infty  {\left[ {\frac{{\hbar ^2 }}{{2\mu }}(u_{n} ')^2  + \left( {\frac{{n(n - 1)\hbar ^2 }}{{2\mu r^2 }} - \frac{{q^2 }}{{4\pi \varepsilon _0 r}}} \right)u_{n} ^2 } \right]} dr
\]
with $c= \int_0^\infty  {u_{n} ^2 dr}$, $u_{n}(r)=rR_{n}$, $u_{n}(0)=0$, $R_{n}^2$ is the probability distribution for the radial location of the electron, $q$ is its charge, $\mu=m_em_p/(m_e+m_p)$ the reduced mass of the system, $\varepsilon _0$ the permitivity of free space and $\hbar$ the Planck's constant divided by $2\pi$. The energies of the system are ruled by the equation $E_{n}=-13.6052\:\textrm{eV}/{n^2}$.

The algorithm was used to find the three first energies of the system. In this case the result found by each run depends strongly in a right choice of $r_N$ such that $u_{n}(r_N)^2\approx 0$. For the ground state $E_1=-13.6052\:\textrm{eV}$ the best found was about eight times the Bohr radius, the average solution found by the algorithm was $E_1=-13.5987\:\textrm{eV}$ with a standard deviation  of $0.028\:\textrm{eV}$. For $n=2$, $E_2=-3.4014\:\textrm{eV}$, the best choice for $r_N$ made was about fifteen times the Bohr radius and  the average solution found by the algorithm was $E_2=-3.42467\:\textrm{eV}$ with standard deviation  of $0.005\:\textrm{eV}$. For $n=3$, $E_3=-1.5117\:\textrm{eV}$, the best choice for $r_N$ made was about twenty five times the Bohr radius, reaching an average solution of $E_3=-1.5103\:\textrm{eV}$ with standard deviation  of $0.001\:\textrm{eV}$. In the three cases $\sigma=0.005\pi$.

\section{CONCLUSIONS AND FUTURE WORK}
The examples shown were chosen with demostrative purposes. Better aproximations for specific cases can be reached increasing $N$ and improving the choice of $\sigma$, with the extra computational effort it implies. Even thought that it was showed the efficiency of the coding to minimize the functionals presented it will be necessary the development of a theory of difficulty to give a more concise explanation of the kind of problems that could be hard to solve using it, like those already existent for binary genes like \emph{deception}\cite{Gold1,Gold2} and \emph{NK landscapes}\cite{Kauffman}. Another useful future development will be that of general ways of handling problems with constraints. An important potential application of the kind of genetic algorithm presented would be in those cases where there are not analytic solutions available, like in many quantum systems.

\begin{acknowledgments}
I wish to acknowledge Dr. Hugo Alarc\'on for introducing me to the idea of solving variational problems using genetic algorithms.
\end{acknowledgments}

\bibliography{gavar}

\end{document}